\documentstyle[aps,prl,multicol,epsfig]{revtex}
\begin{document}
\title{Relative dispersion in fully developed turbulence: \\
The Richardson's Law and Intermittency Corrections}
\author{G.~Boffetta$^1$ and I.~M.~Sokolov$^2$}
\address{$^{1}$ Dipartimento di Fisica Generale and INFM,\\
Universit\`a di Torino, \\
Via Pietro Giuria 1, I-10125 Torino, Italy}
\address{$^2$ Theoretische Polymerphysik, Universit$\ddot{a}$t Freiburg, \\
Hermann-Herder Stra$\ss$e 3 D-79104 Freiburg i.Br., Germany}
%\date{\today}
\maketitle

\begin{abstract}
Relative dispersion in fully developed turbulence is investigated
by means of direct numerical simulations. Lagrangian statistics
is found to be compatible with Richardson description although 
small systematic deviations are found. The value of the Richardson 
constant is estimated as $C_2 \simeq 0.55$, in a close agreement with 
recent experimental findings [S. Ott and J. Mann J. Fluid Mech. {\bf 422}, 
207 (2000)]. By means of exit-time statistics it is shown that the deviations 
from Richardson's law are a consequence of Eulerian intermittency. 
The measured Lagrangian scaling exponents require a set of
Eulerian structure function exponents $\zeta _{p}$ which are remarkably 
close to standard ones known for fully developed turbulence. 
\end{abstract}

\begin{multicols}{2}
\noindent 

%%%%%%%%%%%%%%%%%%%%%%%%%%%%%%%%%%%%%%%%%%%%%%%%%%%%%%%%%%%%%%%%%%%%%

%%%%%%%%%%% Introduction

The statistics of two particle dispersion is historically one of the 
first issues which has been quantitatively addressed in the study of 
fully developed turbulence. This was done by Richardson, in a pioneering 
work on the properties of dispersion in the atmosphere in 1926 \cite{R26}, 
15 years before the theoretical development by Kolmogorov and Obukhov 
\cite{MY75}. 
Despite this fact, there are still relatively few experimental studies 
on turbulent Lagrangian dispersion. This is essentially due to the 
difficulties to obtain Lagrangian trajectories in fully developed turbulent 
flow. The first studies where done in geophysical flows (see \cite{MY75} 
for a review) in which
Lagrangian tracers are more easily followed. Recently, the problem was 
approached in laboratory experiments \cite{JPT99,OM00} but the results 
are still not conclusive. Moreover, most of the numerical
studies of relative dispersion rely on kinematic simulations in synthetic 
flows \cite{EM96,FV98}. Direct numerical simulation have been done mostly 
for two-dimensional turbulence \cite{ZB94,BS01}.

The scope of this Letter is to contribute to the understanding of
relative dispersion by means of direct numerical simulations of
three dimensional turbulence. In what follows we show the qualitative validity 
of the Richardson's description, and discuss its limitations as posed by 
Lagrangian intermittency, whose properties will be investigated in detail.

%%%%%%%%%%%%% Richardson

Richardson's original description of relative dispersion is based on a
diffusion equation for the probability density function of pair separation 
$p({\bf r},t)$ which in the isotropic case can be written as 
\begin{equation}
{\frac{\partial p({\bf r},t)}{\partial t}}={\frac{1}{r^{2}}}{\frac{\partial 
}{\partial r}}r^{2}K(r){\frac{\partial p({\bf r},t)}{\partial r}}.
\label{eq:1}
\end{equation}
The turbulent eddy diffusivity was empirically established by Richardson to
follow the ``four-thirds law'' $K(r)\propto r^{4/3}$. This law is
a direct consequence of the small-scale velocity statistics, as was
first recognized by Obukhov \cite{O41}. Thus, for $r$
within the inertial range, the dimensional analysis gives
\begin{equation}
K(r)=k_{0}\varepsilon ^{1/3}r^{4/3},
\label{eq:2}
\end{equation}
where $\varepsilon$ is the mean energy dissipation and $k_{0}$ a 
dimensionless constant. We remark that (\ref{eq:2}) does not imply that 
a finite energy flux is necessary for particle dispersion.
Indeed, particle separation is observed also in pseudo-turbulent 
synthetic Gaussian velocity field \cite{EM96,FV98,BCCV99}.
Using (\ref{eq:2}), the solution of (\ref{eq:1}) for $\delta$-distribution 
initial condition has the well known form 
\begin{equation}
p({\bf r},t)={\frac{A}{(k_{0}t)^{3}\varepsilon }}
\exp \left( -{\frac{9r^{2/3}}{4k_{0}\varepsilon ^{1/3}t}}\right)
\label{eq:3}
\end{equation}
where $A=2187/2240 \pi^{3/2}$ is a normalizing factor. The most important
feature of the Richardson distribution (\ref{eq:3}) is non-Gaussianity
with a very pronounced peak at the origin and rather fat tails.
In the past, alternative distributions have been proposed \cite{B52,K66}.
In particular Batchelor \cite{B52} suggested a Gaussian distribution
as a consequence of a diffusivity which depends only on averaged quantities.
Because the available data is scarce, there is still no general
consensus on the real form of pair separation pdf. Recent experimental
works \cite{JPT99,OM00} are in favor of (\ref{eq:3}).

The possibility to describe the dispersion process by means of a diffusion
equation is based on two physical assumptions. The
first is that the velocity field is short correlated in time. Indeed, 
in the limit of velocity field $\delta $-correlated in time
(the so-called Kraichnan model of turbulence) the diffusion equation of the
type of Eq.(\ref{eq:1}) becomes exact \cite{K68,FGV01}. The effects of finite
correlation time have been recently discussed in \cite{BS01,S99,SKB00}.

The second assumption, which is one of the points discussed in this Letter,
is that the dispersion process is self-similar in time, i.e. the scaling
exponents of the moments of the separation 
\begin{equation}
R^{2n}(t)\equiv \langle r^{2n}(t)\rangle =
C_{2n}\varepsilon^{n}t^{\alpha_{2n}}
\label{eq:4}
\end{equation}
have the values $\alpha _{2n}=3n/2,$ as following from dimensional analysis.
If this is the case, a single number, such as the Richardson constant
$C_2$ is sufficient to parameterize turbulent dispersion. 
There is still a large uncertainty on the value of $C_{2}$, ranging from 
$O(10^{-2})-O(10^{-1})$ for kinematic simulations \cite{EM96,FV98} 
to $O(1)$ or more in the case of closure predictions \cite{MY75}. 
A recent experimental investigation gives the value $C_{2}=0.5$ 
\cite{OM00}. 
The hypothesis of self-similarity is reasonable with a self-affine
Eulerian velocity, such as in the case of two-dimensional inverse cascade 
turbulence \cite{BS01}. A recent analysis of a kinematic model with synthetic 
velocity field has shown that Lagrangian self-similarity can be broken 
in presence of Eulerian
intermittency. In this case the exponents $\alpha _{n}$ have been found in
agreement with the prediction of a multifractal approach for Lagrangian
statistics. In particular, the second moment of
relative dispersion is not affected by intermittency, i.e. $\alpha _{2}=3$ 
\cite{BCCV99}, essentially because it is proportional to $\varepsilon^{1}$. 
We remind that Lagrangian
intermittency has been observed also in the case of the so-called strong
anomalous diffusion \cite{CMMV99}. Although in that case the mechanism
leading to intermittency is different (there is no scaling invariant flow),
the implication for Lagrangian description is identical, i.e.
the process can not be described by a Fokker-Planck equation like 
(\ref{eq:1}).

\begin{figure}[htb]
\epsfxsize=8truecm
\epsfbox{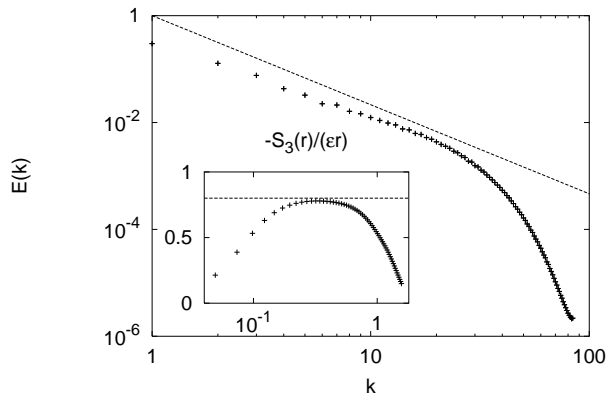}
\caption{ Average energy spectrum $E(k)$. The dashed line has the Kolmogorov
slope $-5/3$. In the inset it is shown the compensated third order
longitudinal structure function. The dashed line represents the ``4/5'' law.}
\label{fig1}
\end{figure}

We now turn to our numerical procedure. The turbulent velocity field is
generated by direct integration of Navier-Stokes equation in a periodic box
of size $L=2\pi $. The integration is done on a parallel computer by means 
of a pseudo-spectral code at resolution $256^{3}$ with 
$Re_{\lambda } \simeq 200$. 
Energy is injected into the flow by keeping the total energy in each of the two
first wavenumber shells constant \cite{CDKS93} and is removed by a 
second-order hyperviscous dissipation.
Time integration is performed with a second-order Runge-Kutta scheme. 
In Figure~\ref{fig1} we plot the energy spectrum which shows a well developed 
Kolmogorov power-law scaling. A small ``bump'' at $k \gtrsim 20$ is the 
signature of a bottleneck effect
\cite{F94} as a consequence of hyperviscosity.
In the inset of Figure~\ref{fig1} we show the third order longitudinal 
structure function $S_{3}(x)=\langle \delta u(x)^{3}\rangle $ compensated 
with the theoretical prediction $S_{3}(x)=-4/5\varepsilon x$ 
\cite{MY75,Frisch95}.

Passive tracer trajectories are obtained by integrating $\dot{{\bf x}}(t)=%
{\bf u}({\bf x}(t),t)$ with the velocity at particle positions obtained by
linear interpolation from the nearest grid points. The reported results are
obtained averaging over a total number of about $3 \times 10^5$ particle pairs 
starting from initial separation $R(0)=L/256$ and over $7$ large scale eddy 
turnover times. 

\begin{figure}[htb]
\epsfxsize=8truecm
\epsfbox{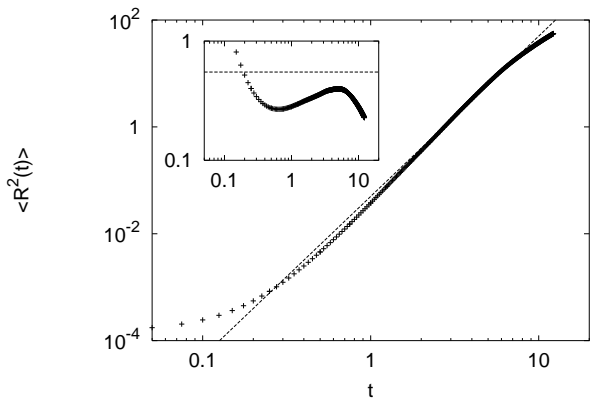}
\caption{Relative dispersion $R^2(t)$ versus time $t$. The dashed line is
the Richardson $t^3$ law. In the inset we show the compensated plot
$R^2(t)/(\varepsilon t^3)$ which should give the Richardson constant $C_2$. 
Because of the strong oscillation, a precise estimation of $C_2$
is difficult.}
\label{fig2}
\end{figure}

In Figure~\ref{fig2} we plot the second moment of relative dispersion
$R^{2}(t)$. The Richardson $t^{3}$ law (\ref{eq:4}) is clearly observable
although systematic deviations are detectable, in particular in the
compensated plot. These deviations, observed also in kinematic simulations 
\cite{BCCV99} and in two-dimensional turbulence \cite{BS01}, are 
due to finite size effects. Consider a series of pair dispersion
experiments, in each of which a couple of particles is released at time $t=0$
at initial separation $R(0)$. At a fixed time $t$ one performs an average 
over all realizations and computes $R^2(t)$.
For $t$ small $R^2(t)$ is dominated by the initial distance, so that the
$R^2(t)$-curve flattens. For large times
some pairs might have reached a separation larger than the integral 
scale and thus show normal (not Richardson) diffusion, so that the 
$R^2(t)$-dependence flattens again.
Under these conditions, a precise determination of the exponents and 
coefficients in (\ref{eq:4}), in particular the Richardson constant $C_{2}$, 
is very difficult.

\begin{figure}[htb]
\epsfxsize=8truecm
\epsfbox{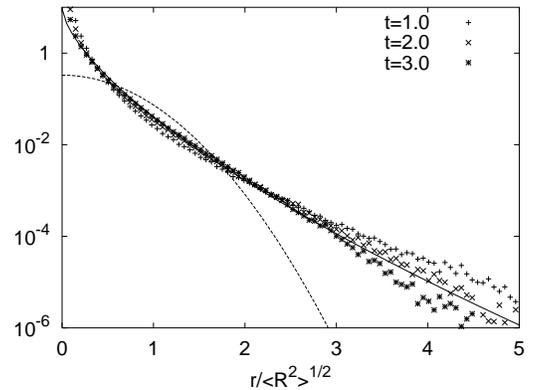}
\caption{Probability distribution function of relative separations at three
different times. The continuous line is the Richardson prediction 
(\ref{eq:3}), the dashed line is the Gaussian distribution proposed by 
Batchelor.}
\label{fig3}
\end{figure}

The distribution of relative separations is plotted in Figure~\ref{fig3} for
three different times. The form of the pdf is very close to the
Richardson prediction (\ref{eq:3}) and excludes other distributions.
Our result is the first direct numerical evidence of the substantial 
validity of Richardson's equation and gives
support to recent experimental findings \cite{OM00}. A closer inspection of
Figure~\ref{fig3} reveals however that the self-similar evolution predicted by 
(\ref{eq:1}) is not exact. Again, the deviations from the distribution 
(\ref{eq:3}) are mostly due to finite Reynolds effects: because of the
large tails, a large fraction of particles exits the inertial range after 
a very short time.

To overcome these difficulties in Lagrangian statistics, an alternative
approach based on {\it doubling time} (or exit time) statistics has been
recently proposed \cite{BCCV99,ABCCV97}. Given a set of thresholds 
$R_{n}=\rho ^{n}R(0)$ within the inertial range, one computes the ``doubling
time'' $T_{\rho }(R_{n})$ defined as the time it takes for the particle pair
separation to grow from threshold $R_{n}$ to the next one $R_{n+1}$.
Averages are then performed over many dispersion experiments, i.e., particle
pairs. The outstanding advantage
of averaging at fixed scale separation, as opposite to a fixed
time, is that it removes crossover effects since all sampled particle pairs
belong to the same scales. In the simulations presented here, the value 
$\rho =1.2$ is used.

\begin{figure}[htb]
\epsfxsize=8truecm
\epsfbox{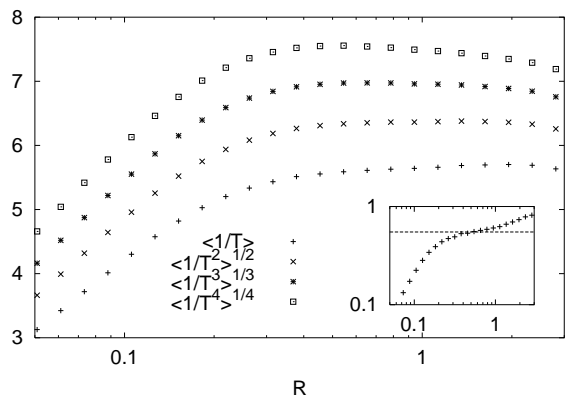}
\caption{First moments of the inverse doubling time $\langle (1/T(R))^p
\rangle$ compensated with Kolmogorov scaling $R^{-2p/3}$. Deviations from
dimensional compensation are evident, in particular for $p=4$. In the inset
we plot the compensated mean doubling time according to (\ref{eq:6}) 
together with the estimate corresponding to $C_2 \simeq 0.55$.}
\label{fig4}
\end{figure}

Let us first show how doubling time analysis can be used for estimating the
Richardson constant $C_{2}$. Neglecting intermittency, the mean doubling 
time can be obtained from the first-passage problem for the Richardson 
diffusion equation (\ref{eq:1}) as 
\cite{BS01} 
\begin{equation}
\langle T_{\rho }(R)\rangle ={\frac{\rho ^{2/3}-1}{2k_{0}
\varepsilon^{1/3}\rho ^{2/3}}}R^{2/3}.
\label{eq:5}
\end{equation}
From (\ref{eq:3}) and (\ref{eq:4}) one has $C_{2}={\frac{1144}{81}}k_{0}^{3}$.
Comparison with (\ref{eq:5}) gives 
\begin{equation}
C_{2}={\frac{143}{81}}{\frac{(\rho ^{2/3}-1)^{3}}{\rho ^{2}}}
{\frac{R^{2}}{\varepsilon \langle T_{\rho }\rangle ^{3}}}.
\label{eq:6}
\end{equation}
In the inset of Figure~\ref{fig4} we plot expression (\ref{eq:6}) which
gives directly the value of $C_{2}$. Although the compensation is not
perfect, it is possible to estimate the Richardson constant with much 
better accuracy than from the direct analysis of Figure~\ref{fig2}.
The resulting value, $C_{2}=0.55\pm 0.1$, is
remarkably close to the recent experimental finding \cite{OM00}.
The non perfect compensation is the consequence of intermittency.

Let us now discuss the issue of intermittency in more detail and
concentrate on the behavior of the moments of inverse doubling times,
$\langle \left( 1/T_{\rho }(R)\right)^{p}\rangle$. 
We expect for doubling time statistics a power-law behavior 
\begin{equation}
\langle \left( {\frac{1}{T_{\rho }(R)}}\right)^{p}\rangle 
\sim R^{\beta_{p}}
\label{eq:7}
\end{equation}
with exponents $\beta _{p}$ connected to the exponents $\alpha _{n}$ in 
(\ref{eq:4}). Negative moments of doubling time are dominated by pairs which
separate fast; this corresponds to positive moments of relative separation.
Kolmogorov scaling, based on the dimensional analysis, gives 
$\langle \left( 1/T_{\rho }(R)\right)^{p}\rangle \sim 
\varepsilon ^{p/3}R^{-2p/3}$ so that $\beta _{p}=-2p/3$. Intermittency can be
taken into account by using the simple dimensional estimate for the
doubling time, $T(R)\sim R/\delta u(R)$ which gives 
\begin{equation}
\beta _{p}=\zeta _{p}-p ,
\label{eq:8}
\end{equation}
where $\zeta_{p}$ are the scaling exponents of the longitudinal structure
functions. As a consequence of the Kolmogorov ``4/5'' law, $\zeta_{3}=1$ 
\cite{Frisch95} and the doubling time exponent not affected by intermittency
is $\beta _{3}=-2$ (again, the quantity not affected by intermittency depends 
on the first power of $\varepsilon$) \cite{note}.

In Figure~\ref{fig4} we plot the first moments of inverse doubling time (\ref
{eq:7}) compensated with the Kolmogorov scaling $R^{-2p/3}$. The quality of
the scaling is remarkable, especially if compared with the standard
statistics of Figure~\ref{fig2}. This allows to detect small
deviations from dimensional scaling. 
Indeed, a closer inspection of Figure~\ref{fig4} reveals that the
compensation is not perfect, the deviation being more evident for higher
moments; this indicates the existence of Lagrangian
intermittency.

\begin{figure}[htb]
\epsfxsize=8truecm
\epsfbox{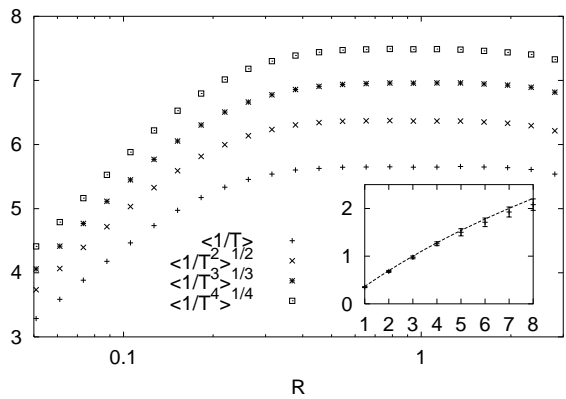}
\caption{First moments of the inverse doubling time $\langle (1/T(R))^p
\rangle$ compensated with best fit exponent $\beta_{p}$. Observe the
improvement in the compensation with respect to Figure~\ref{fig4}. In the
inset we plot the structure function exponents estimated from $%
\zeta_p=p+\beta_p$. The line is the She-Leveque parameterization.}
\label{fig5}
\end{figure}

Figure~\ref{fig5} shows some moments of the inverse doubling time, now
compensated with best fit exponents $\beta _{p}$. The improvement with
respect to Figure~\ref{fig4} demonstrates that the exponents in (\ref{eq:7})
are corrected in comparison to dimensional prediction. From the doubling
time exponent $\beta _{p}$ we can obtain the Eulerian exponents $\zeta_{p}$
by inverting (\ref{eq:8}). The result shown in the inset of Figure~\ref{fig4}
gives a set of exponents $\zeta_{p}$ which are remarkably close to
``standard'' structure function exponents in fully developed turbulence. We
stress that, at the present resolution, the scaling of the Eulerian
structure function is rather poor, thus a precise determination of $\zeta _{p}$ 
is possible only using indirect analysis, such as the
ESS technique \cite{BCTBMS93}.

Let us summarize our findings. We have performed direct numerical
simulations of a three-dimensional turbulent flow and concentrated on the
problem of particles' dispersion. The overall dispersion
behavior is well-described by the Richardson's equation, although
some deviations (mostly caused by the finite-Reynolds nature of the
simulations) are evident. The use of fixed-scale statistics (doubling-time
distribution) instead of fixed-time ones removes to a large extent these
restrictions, and gives a possibility to evaluation the
Richardson's constant very accurately. Its value is $C_{2}\simeq 0.55$, 
in a close
agreement with recent experimental findings. The discussion of the inverse
moments of the doubling-time distributions unveils the role of Lagrangian
intermittency in the two-particle dispersion. The values of
the Lagrangian scaling exponents are connected with the Eulerian structure
function exponents $\zeta_{p}$. The values of $\zeta_{p}$ obtained from
the separation statistics are remarkably close to standard ones, known for
fully developed turbulence. In the next future it will be probably possible
to have experimental Lagrangian trajectories in high Reynolds number flows 
\cite{LVCAB01}. It would be extremely interesting to check our findings in
real fluid turbulence.

\begin{acknowledgments}
We thank A.~Celani and M.~Cencini for useful comments and discussions.
We acknowledge the allocation of computer resources
from INFM Progetto Calcolo Parallelo.
\end{acknowledgments}                                                           

%%%%%%%%%%%%%%%%%%%%%%%%%%%%%%%%%%%%%%%%%%%%%%%%%%%%%%%%%%%%%%%%%%%%%

%%%%%%%%%%%%%%%%%%%%%%%%%%%%%%%%%%%%%%%%%%%%%%%%%%%%%%%%%%%%%
%\newpage
\narrowtext

\newpage

%%%%%%%%%%%%%%%%%%%%%%%%%%%%%%%%%%%%%%%%%%%%%%%%%%%%%%%%%%%%%%%%%%%%
\end{multicols}

\begin{references}
\bibitem{note} It may seem that the $-3$rd moment of $T_{\rho }(R)$, not 
affected by intermittency, may give a better estimate for $C_{2}$ than 
its first moment, Eq.(5). This is not the case, due to a remarkably 
strong $\rho $-dependence of $\left\langle (1/T_{\rho }(R))^{3}\right\rangle $ 
and to its strong sensitivity to temporal correlations. The corresponding 
results will be discussed elsewhere.
\end{references}
\end{document}